\documentclass[submission,copyright,creativecommons]{eptcs}
\providecommand{\event}{TAV-WEB 2010} 
\usepackage{breakurl}             
\usepackage[ruled,linesnumbered]{algorithm2e}
\usepackage{graphicx}
\usepackage{multirow}
\usepackage{float}
\usepackage{amsmath}
\usepackage{enumerate}
\usepackage{mdwlist}
\usepackage{tweaklist}

\newcommand{\bit}{\begin{itemize*}}
\newcommand{\eit}{\end{itemize*}}

\renewcommand{\enumhook}{\setlength{\topsep}{0pt}\\
  \setlength{\itemsep}{0pt}}

\newcommand{\comment}[1]{}

\graphicspath{{figs/}}
\DeclareGraphicsExtensions{.ps,.eps,.epsi}

\title{Structural Learning of Attack Vectors for Generating Mutated XSS Attacks}
\author{$^{1}$Yi-Hsun Wang  \qquad $^{1}$Ching-Hao Mao \qquad $^{1,2}$Hahn-Ming Lee
\institute{$^{1}$Department of Computer Science and Information Engineering\\ National Taiwan University of Science and Technology\\
Taipei, Taiwan}
\institute{$^{2}$Institute of Information Science\\ Academia Sinica\\
Taipei, Taiwan}
\email{\{m9715053,d9415004,hmlee\}@mail.ntust.edu.tw}
}
\def\titlerunning{Structural Learning of Attack Vectors for Generating Mutated XSS Attacks}
\def\authorrunning{Y.H. Wang, C.H. Mao \& H.M. Lee}
\begin{document}
\maketitle
\begin{abstract}
Web applications suffer from cross-site scripting (XSS) attacks that resulting from incomplete or incorrect input sanitization. Learning the
structure of attack vectors could enrich the variety of manifestations in generated XSS attacks. In
this study, we focus on generating more threatening XSS attacks for the state-of-the-art detection
approaches that can find potential XSS vulnerabilities in Web applications, and propose a mechanism
for structural learning of attack vectors with the aim of generating mutated XSS attacks in
a fully automatic way. Mutated XSS attack generation depends on the analysis of attack vectors
and the structural learning mechanism. For the kernel of the learning mechanism, we use a Hidden
Markov model (HMM) as the structure of the attack vector model to capture the implicit manner of
the attack vector, and this manner is benefited from the syntax meanings that are labeled by the proposed
tokenizing mechanism. Bayes¡¦ theorem is used to determine the number of hidden states in the
model for generalizing the structure model. The paper has the contributions are as following: (1)
automatically learn the structure of attack vectors from practical data analysis to modeling a structure model of attack vectors, (2) mimic the manners and the elements of attack vectors to extend the ability of testing tool for identifying XSS vulnerabilities, (3) be helpful to verify the flaws of blacklist sanitization procedures of Web applications. We evaluated the
proposed mechanism by Burp Intruder with a dataset collected from
public XSS archives. The results shows that mutated XSS attack
generation can identify potential vulnerabilities.
\end{abstract}

\section{Introduction}\label{sec:intro}
According to the Open Web Application Security Project
(OWASP)~\cite{OWASPTop10}, cross-site scripting (XSS) is already the
one of top two vulnerabilities in Web applications. The XSS
vulnerabilities are due to using mistrusted data from the out of Web
application as the partial contents of HTML page's output and this attack
may result in information disclosure~\cite{CookieStealing}.
Furthermore, the implementation of HTML interpreters embedded in
browsers incompletely comply with specifications or provide
extra functionalities (e.g., browser-specific HTML tags, attributes,
and events). In the above situations, Web application programmers
difficultly sanitize the input message for preventing the XSS
vulnerabilities by uniform XSS patterns or detection rules.

In general, an XSS attack can be separated into
the attack vector and the attack body~\cite{Fogla06polymorphicblending}.
An attack body is the main code for executing the intention (e.g.,
it can invoke JavaScript interpreter) after exploiting a vulnerability successfully, and it is often applied by
obfuscation techniques beyond the detections. An attack
vector~\cite{AttackVectorAtSearchSecurity, Hansman200531} is the
medium for introducing the attack body. If imagining a XSS exploit as a missile, the attack vector is like the guided device of the missile, and the attack body is like the warhead of the missile. Hence, an attacker can promote the attack body to be
interpreted for malicious intension by using the right or efficient attack vectors.

\begin{figure}
\center
\includegraphics[scale = 0.5]{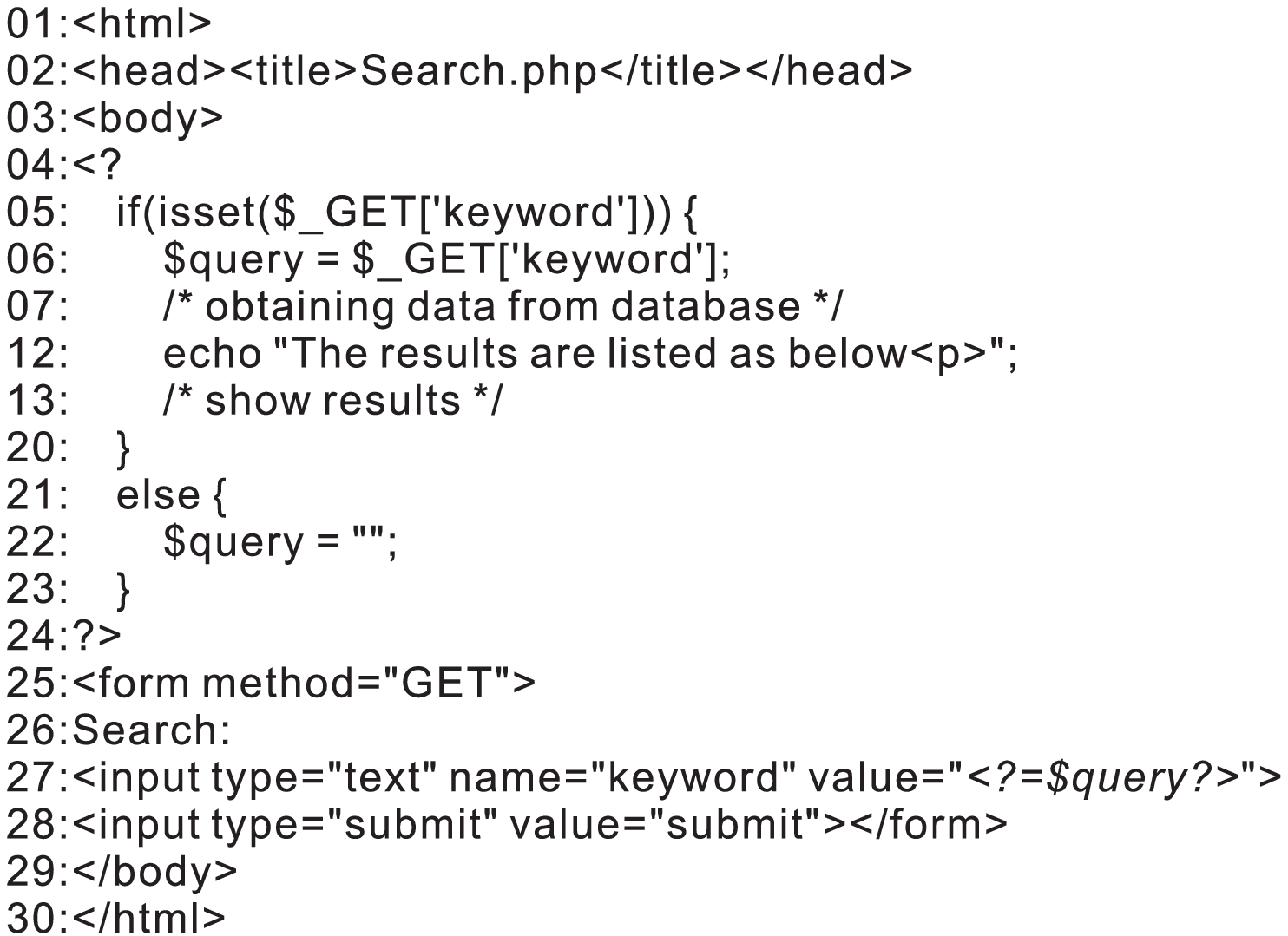}
\caption[Motivation Example from PHP code]{A motivation example. There is a XSS vulnerability at line 27 of this page. Someone can input a string prefixed double quote by variable ``keyword'' to trigger the vulnerability. When the form is submitted, the input string is assessed whether assigning a value at line 5. If true, the input string will be assigned to variable ``query''. Finally, the variable ``query'' becomes the value of attribute ``value'' in the input element, and it is included between double quotes.}
\label{fig:Search}
\end{figure}

In Figure~\ref{fig:Search}, we depict the relationship between
the attack vector and the attack body in the example code. It is
noticeable that there is a XSS vulnerability at line 27 of this
page. The vulnerability is the result of the input being used as the value of attribute ``value'' in the input element.
Someone can input a string prefixed double quote like, ``\emph{''$<$script$>$alert(123)$<$/script$>$}'', through variable
``keyword'' to trigger the vulnerability.
However, the following attack may not work, ``\emph{$<$script$>$alert(123)$<$/script$>$}''.
It is clear that double quote is the critical character to introduce
the attack body (i.e.,
``\emph{$<$script$>$alert(123)$<$/script$>$}''), and the attack
vector is double quote in here. The string
``\emph{"$><$script$>$alert(123)$<$/script$><$a name=$"$}'' is even
better, as it is seamlessly embedded in the page.

Although a lot of previous studies focus on creating attack vectors
to identify XSS vulnerabilities, the involved attack vectors are
either generated based on path constraints or equipped a set of attack payloads.
However, when reviewing
XSS cheat sheet~\cite{XSSCheatSheet} and public XSS
exploits~\cite{xssed}, it is quite obvious that attackers often
use special symbols and HTML elements for crafting XSS attacks in some manners. Thus, we
attempt to propose a structural learning mechanism of the attack vector
for generating mutated XSS attacks.

The proposed mechanism presents an automated technique for
generating mutated XSS attacks to test XSS vulnerabilities in Web
applications. First, the system automatically crawls public XSS attacks existed in URLs, and extracts the elements of XSS attacks. Next, the structural learning module builds a structure model based on a hidden Markov model (HMM) while generalizing the structure of the model by Bayes' Theorem. Finally, according to the structure model, the system generates mutated XSS attacks based on Viterbi algorithm to enlarge the detection capability of the testing tool.

The challenge in generating mutated XSS attacks is how to compose
the right element in the right position of the structure to exploit the specified vulnerabilities. If the attack vector can not pass through the internal procedures (e.g., sanitization function, string manipulation function) of the Web application, the XSS attack maybe fail. There are many possible ways to invoke the JavaScript
interpreter according to W3C and browser
specifications. In such situations, it is impractical for an expert to spend much time identifying a vulnerability.
Hence, learning the manners and elements from the known XSS attacks is helpful for mutated attacks generation.

The goal of the proposed technique, different from previous works of Web applications testing, is aimed
at learning the elements and the implicit structure which are presented within the XSS
attacks. Thereby, learning from public XSS exploits can benefit from the common weaknesses of Web applications
that cannot be handled well, and increasing the
probability of identifying XSS vulnerabilities in black or white-box
testing. In the other hand, learning the structure of attack vectors
could enrich the manifestations of XSS attacks.
In summary, the contributions of our proposed approach are as following:
\begin{enumerate}
\item can automatically learn the structure of attack vectors from practical data analysis to modeling a structure model of attack vectors.
\item can mimic the manners and the elements of attack vectors to extend the ability of testing tool for identifying XSS vulnerabilities.
\item can be helpful to verify the flaws of blacklist sanitization procedures of Web applications.
\end{enumerate}

The rest of this paper is organized as follows: First, the proposed approach for generating mutated XSS attacks in
Section~\ref{sec:frame}. The experiment design and the evaluation are given in Section~\ref{sec:exp}. After that, the related
work will be introduced in Section~\ref{sec:past}. Finally, the conclusion will be briefed in Section~\ref{sec:conc}. 
\section{Generation of Mutated XSS Attacks}\label{sec:frame}
The goal of our proposed approach is to learn the structure
of attack vectors and generate mutated XSS attacks. The proposed mechanism consists of three components, the attack vector tokenizer, attack vector
inducer, mutated attack generator, and one attack vector profile.

\begin{figure}
\center
\includegraphics[scale=0.4, trim=0 80 0 80, clip]{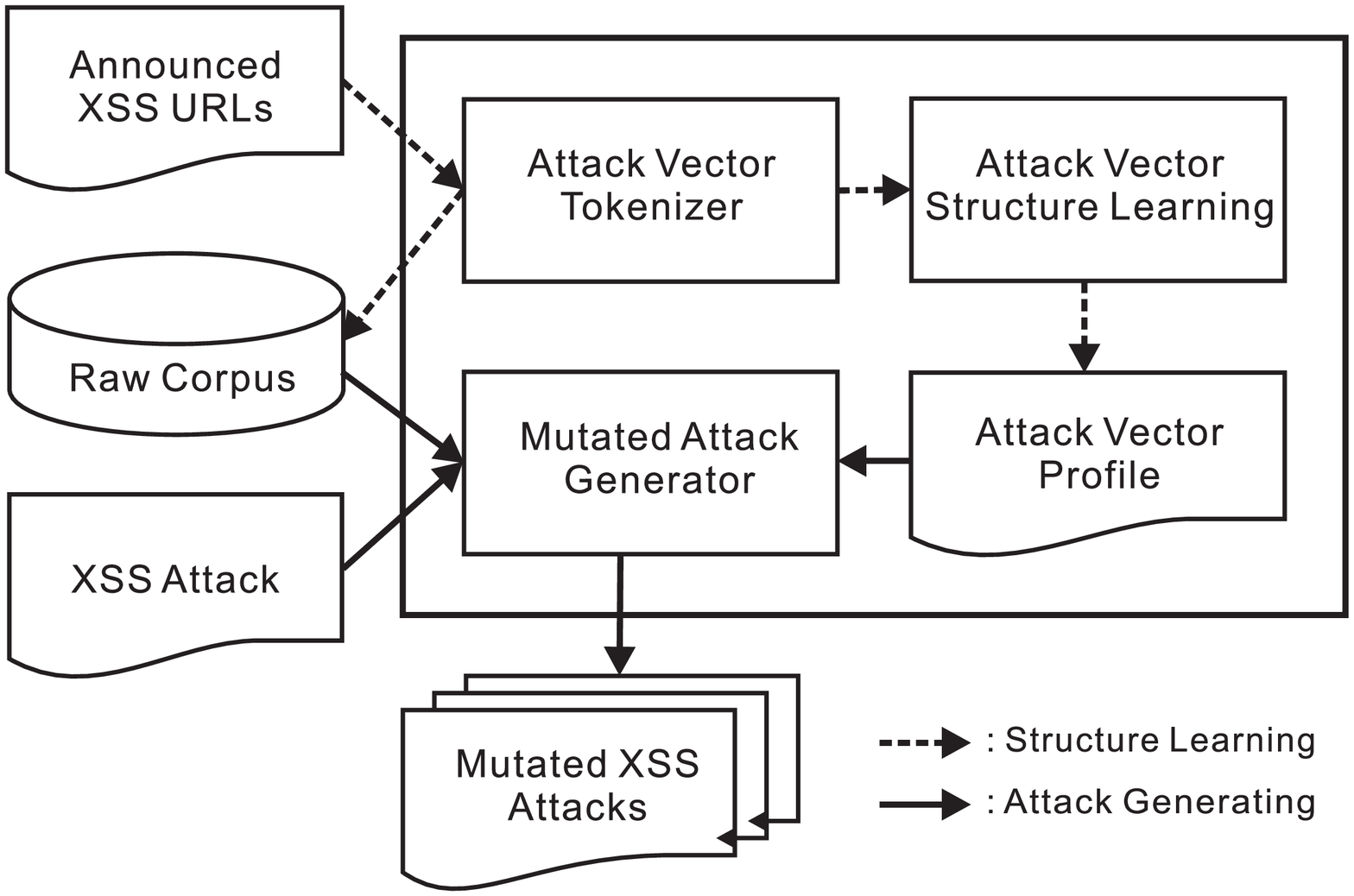}
\caption[System Architecture]{Generation of Mutated XSS Attacks.
In the structure learning phase,
the attack vector tokenizer crawls the XSS URLs as input, and produces a profile of the attack vector as an output. In the attack generating phase, the mutated attack generator estimates the most possible state sequence of referred XSS attack \comment{based on Viterbi~\cite{1967Viterbi} algorithm} based on the attack vector profile, and acquires the relative elements from raw corpus for generating mutated XSS attacks.}
\label{fig:SA}
\end{figure}

Mutation involves two phases, a structure learning phase and an attack generating
phase. In the structure learning phase, the attack vector tokenizer
attempts to extract the elements of attack vectors from known XSS
URLs. Next, the attack vector inducer profiles the sequential relations among the attack elements and the attack vector profile records these sequential relations for mutating XSS attacks. In the attack generation phase, the mutation mechanism based on the XSS
attacks, and transforms them into a sequence of tokens by passing it
through the attack vector tokenizer. Subsequently, the mutated attack
generator based on the structure of referred XSS attacks and the attack vector profile to generate mutated XSS attacks from raw corpus.
Thus, the proposed approach is an automatic way from learning the structure of
attack vectors to generating mutated XSS attacks. An overview of the structural learning mechanism is depicted in Figure~\ref{fig:SA}.

\subsection{Attack Vector Tokenizer}
The attack vector tokenizer could be regarded as preprocessing
module with three functionalities, including decoding,
identification and tokenization. The task of the attack vector tokenizer is automatic collecting the announced XSS URLs and disassembling them from public Web sites (e.g.,~\cite{xssed}) or information security organizations. The attack vector tokenizer has a XSS
attack locator and a token extractor for handling XSS URLs. First, the XSS attack locator crawls the XSS URLs and seeks the parameters
in which the XSS attack existed. Next, the token extractor abstracts the XSS attacks and outputs the tokens for subsequent
procedures. Thus, the attack vector tokenizer is able to identify an XSS attack from a XSS URL and extract the
elements of attack vectors. These components of the attack vector
tokenizer are described in greater detail in the following subsections.

\subsubsection{XSS Attack Locator and Token Extractor}
The XSS attack locator is responsible for identifying an XSS attack from a XSS URL. Malicious users are familiar with obfuscating their
crafted XSS attacks to evade the detection mechanism. HTML entity
codes, URL encoding, Base64 and double encoding are frequently observed obfuscation techniques. Consequently, if the XSS attack locator
directly processes these obfuscated XSS URLs, not only it easily fails to identify the XSS attack, but the token extractor will also lose a lot of
structural information about the attack vector. To obtain the
original structure of the attack vector, the XSS attack locator handles the decoding of HTML symbol entities and URL encoding, besides system control characters.

After decoding, the XSS attack locator examines the entire
XSS URL to identify where the XSS attack is. According to URI
syntax~\cite{2005:URI:RFC3986}, essentially three delimiters are considered in splitting the entire URL to recognize the position from the
values. The term "value" in here is the value of a parameter in a URN. The XSS attack
locator complies with the delimiters to obtain one or more
values in a XSS URL, and estimates the most possible value where the XSS attack existed by six weighted features.

The token extractor is responsible for transforming a XSS attack into a
sequence of tokens. The types of token are as follows: start\_tag, attribute,
the value of an attribute, plain text, end\_tag, and comment. When the
token extractor receives the XSS attack, it disassembles the XSS attack into a sequence of tokens and saves the original substrings
to the raw corpus. The raw corpus records the mapping information
between the type of token and the original substring.

\subsection{Structural Learning of Attack Vectors}\label{sec:inducer}
In the real XSS attacks, they are often mutated with a few attack
vectors for discovering increasingly the XSS vulnerabilities. In
order to mimic the manners and elements from attack vectors, we
propose a Structural Learning of Attack Vectors (SLAV) mechanism
that can profile attack vectors in an automatic way. From learning,
the SLAV attempts to learn the sequential relations among the
elements of attack vectors and these relations can be regarded as a
type of implicit regular grammar that defines all possible
combination of elements.

In the SLAV, we apply probabilistic models for discovering the
implicit meaning behind observed elements, and the model is regarded
as the structure model of attack vectors. Briefly, the structural
learning procedure is comprised of three steps as follows.
\begin{enumerate}
\item Incorporate each token sequence into the current structure model.
\item Adjust the state topology of the model to match all of the data.
\item Output a profile of the structure model when finishing to incorporating all data.
\end{enumerate}
\noindent These steps are described in full detail in the following subsections.

\subsubsection{Attack Vector Structure Learning}\label{secsec:BFHM}
To infer the general forms of attack vectors from token sequences,
the attack vector structure learning (or saying AVSL) intends to
model stochastic sequences within an finite states based on HMM.
Also, this structure model of attack vectors is adjusted until it is
general enough to describe all XSS attacks by Bayes' Theorem. Thus,
the entire structure model can be viewed as a probabilistic
automaton~\cite{2005_PR}. In a probabilistic automaton, each
production rule is assigned a probability, and it implies that some
attack vectors are more likely to be generated than others. In a
HMM, each state represents a non-terminal symbol, each observation
represents a terminal symbol in grammar, and all paths that are
possible routes from the starting state to the final state denote
all possible production rules with non-zero probability (i.e., the
product of initial probability, transition probability and emission
probability). The HMM benefits the model capturing the sequential
relations of tokens and obtaining the implicit grammatical manners
for generating mutated XSS attacks.

For generalization, the Bayes' Theorem can be expressed by the following equation.
\begin{equation}
P(M|X) = \frac{P(M)P(X|M)}{P(X)}
\end{equation}
Here, $M$ is the structure model, and $X$ is a set of training token sequences.
Both $X$ and $P(X)$ are constant and can be ignored in
the computation. In order to maximize the a posteriori probability
$P(M|X)$, the product of $P(M)$ and $P(X|M)$ requires to be
maximized. $P(M)$ is the prior probability of the model and is
assigned by the Dirichlet distribution. $P(X|M)$ is the probability of a set of training token sequence given the model and can be calculated for a model topology
(i.e., state transition and emission probabilities) with a various probabilities and performances for a
fixed training tokens. Thus, Bayes' Theorem is used to find the
balance between $P(M)$ and $P(X|M)$ and to decide the
appropriately general model for representing the form of XSS attacks without bias.

The goal of the AVSL is building a structure model with the highest generalization for describing the relations of all elements.
Model-building starts with an empty structure model without any tokens. In step 1, when a token sequence coming, it is initially
modeled as a HMM~\cite{1986:HMM} and incorporate into the current structure model. Then, in the current structure model, all of the possible pairs of states are merged and examined by Bayes' Theorem for obtaining a merged model with the maximum posteriori probability in step 2. In step 3, if the a posteriori probability of the merged model is less than that of the current model, then the current structure model is replaced with the merged model. After repeating above steps to each token sequence, a profile of the current structure model is published. For details of the complete process, the readers are referred to~\cite{1993_NIPS} and~\cite{TR-Stolcke94best-firstmodel}.

\subsubsection{Attack Vector Profile}
Once the structure model has been built, an attack vector profile is
published. An attack vector profile that describes all parameters of
the structure model. These parameters contain a set of hidden
states, a set of tokens, the initial probability of each state, the
transition probabilities that record the likelihood of transitioning
from one state to next, and the emission probabilities that record
the probability distribution over the possible tokens in each state.
The existing of the attack vector profile provides the information
of the structure model, that is followed to generate mutated XSS
attacks.

\subsection{Mutated Attack Generator}\label{subsec:Mutate}
The mutated attack generator attempts to simulate the structure of
referred XSS attack by evaluating the possible state sequence in the
structure model, and applying the original elements from raw corpus.
In the attack generation phase, the procedure starts with tokenizing
the referred XSS attack as a sequence of tokens. Subsequently, the
mutated attack generator estimates the token sequence and seeks the
maximum-likelihood state sequence by the Viterbi
algorithm~\cite{1967Viterbi} according to the attack vector profile.

The Viterbi algorithm finds the maximum-likelihood sequence of
hidden states--also called the Viterbi path--for a given sequence of
tokens and profile of HMM. For purposes of generation, the Viterbi
path can be interpreted as the context in which a given XSS attack
takes place. Therefore, the mutated attack generator is able to
acquire the structure of attack vectors for generating a variety of
XSS attacks. In addition, functions embedded in the mutated attack
generator include mutating attacks with obfuscation techniques, such
as URL encoding and HTML entity codes.

\section{Experiment and Results}\label{sec:exp}
In this section, we evaluated the proposed approach by the
experiment that is designed for generating mutated attacks to
penetrate the vulnerabilities of PHP Web applications. The used dataset and the experiment design concept are mentioned in
Subsection~\ref{subsec:data}. The evaluation metrics are described for
evaluating the performance of proposed approach in the aspect of
vulnerability discovering are given in Subsection~\ref{subsec:metrics}.
The experiment results and case studies are given in Subsection~\ref{subsec:result}.
Finally, we discuss the proposed approach in Subsection~\ref{subsec:analyze}

\subsection{Experiment Design and Dataset}\label{subsec:data}
The goal of experiment intended to measure the effectiveness of
mutated XSS attacks in identifying XSS vulnerabilities in PHP Web
applications. It is noticeable that we neither claim to completely
cover the space of possible variations of an attack, nor states that
we guarantee that all the possible mutated attack are successful.
Nevertheless, we aim at providing an effective mechanism for the
structural learning of attack vectors to test the Web application.
The experiment was designed to answer following research questions:
\begin{itemize}
\item How is the effectiveness of the mutated XSS attack generation to find out the existed vulnerabilities?
\item What the manifestation of XSS attack vectors trigger potential vulnerabilities?
\end{itemize}

The configuration procedures for each target program are listed as
following. First, we set up an environment for the requirements of
each target program (e.g., essential configurations, database tables
and so on). Second, we built the structural model of attack vectors
from the public dataset and used the model to generate XSS attacks
based on the attack information (i.e., the relative parameters and
the attack body). Third, the Burp Intruder~\cite{Burp}, a famous
tool for automating attacks against Web applications and allowing
for importing customized attack payloads, run the XSS attacks
generated by proposed approach against the target Web applications.
Finally, we investigated manually to check the results of XSS
attacks that were reported as successful attacks whether really
trigger attack bodies.

For the evaluation, the dataset (denoted as XSSed10) was collected
from XSSed project\footnote{XSSed project: http://xssed.com}, and
the duration of the collection was between 2009/05/16 and
2010/03/13. Totally, the XSSed10 had 3019 XSS attacks for this work
and several obvious characteristics were in XSSed10. For example,
one or more special symbols were outside of tags (e.g.,
"$>$"$><$script$>$) and encoding techniques were in favor.

In the structure model, we obtained totally 11 states, 1031 tokens
and 192 unique token sequences from the XSSed10. \comment{
\begin{figure}
\begin{center}
\includegraphics[scale=0.3, trim=0 0 0 0, clip]{xssed.com.ps}
\caption[The structural model learned from XSSed10]{The structural model learned from XSSed10. The node 1 was the starting node as well as ending node. Each node represented certain type of token, and the value in the node was the emission probability. The per value at per edge denoted the transition probability between each state in the figure.}
\label{fig:mymodel}
\end{center}
\end{figure}
}

The testing target were the two PHP open-source Web applications
listed in Table~\ref{table:php program}. The
``schoolmate''~\cite{schoolmate} is a Web solution for assisting the
elementary, junior and senior high school to provide the
administration of school affairs. Another testing target naming
''Webchess''~\cite{webchess} is an online chess game. There have
been several identified vulnerabilities in these two open-source PHP
Web applications, and hence, they were suitable to be used for
evaluating the effectiveness of mutated XSS attacks whether they
could trigger the practical vulnerabilities or not.

\begin{table}
\begin{center}
\caption[The characteristics of target Web applications]{The characteristics
of Web applications (The \emph{Vulnerabilities} column lists the
number of XSS vulnerabilities mentioned in~\cite{2009:Ardilla}. The \emph{Downloads} column lists
downloads counted by Sourceforge~\cite{SourceForge} at June 6th 2010.)}
\label{table:php program}
\begin{tabular}{@{}ccc p{15cm}@{}}
\hline \textbf{Program} (Version) &\textbf{Vulnerabilities} &\textbf{Downloads} \\
\hline
webchess (0.9.0)~\cite{webchess}     & 13 & 41257 \\
schoolmate (1.5.4)~\cite{schoolmate} & 18 & 6900 \\
\hline
\end{tabular}
\end{center}
\end{table}

The referred information of XSS attacks were
from~\cite{2009:Ardilla}, and they were base XSS attacks for
mutation. There were many XSS vulnerabilities identified by these
XSS attacks in schoolmate and webchess. Before the system generated
mutated XSS attacks, the base XSS attack was tokenized as a token
sequence. Then the system obtained a sequence of hidden state
according to the token sequence by Viterbi algorithm. Finally,
mutated XSS attacks were generated by composing the elements of raw
corpus. Especially, each mutated XSS attack contained attack body,
$<$script$>$alert(123)$<$/script$>$ for test oracle.

\subsection{Evaluation Metrics}\label{subsec:metrics}
Because the target programs have been examined by previous
researches, we regarded to the effectiveness of our system for
assisting vulnerability scanner in attack string generation by two
metrics, that is, the false positive rate (or saying FP rate) and
the recall rate (or saying Recall). In here, the true positive means
the attack can invoke the JavaScript interpreter. The false positive
means the reported successful attack, but it cannot invoke the
JavaScript interpreter. The summary of true positives and false
positives are the total generated attacks. Hence, the FP rate
denotes the ratio of the number of failed testing attacks in to the
number of successful testing attacks in report, shown as
Equation~\ref{eq:FP}. This indication shows the level of threat of
the mutated XSS attacks, and the more lower FP rate denotes that the
set of mutated XSS attacks are more threatening to a target Web
application.

\begin{equation}
\label{eq:FP}
FP = \frac{\#~of~real~failed~attacks~in~successful~attacks}{\#~of~whole~reported~successful~attacks}
\end{equation}

Recall rate is used to measure the ability of mutated XSS attacks
about identifying vulnerability in Web applications. In here, the
numbers of total vulnerabilities were collected
from~\cite{2009:Ardilla} with 18 and 13 vulnerabilities
respectively. So, the Recall rate is \comment{the ratio of the
number of found vulnerabilities to the number of whole
vulnerabilities in target Web application, } shown as
Equation~\ref{eq:Recall}. The higher Recall rate means that the set
of mutated XSS attacks are with much more capability to identify XSS
vulnerabilities.

\begin{equation}
\label{eq:Recall}
Recall = \frac{\#~of~found~vulnerabilities}{\#~of~whole~vulnerabilities~in~target~Web~application}
\end{equation}

\subsection{Numerical Results and Case Studies}\label{subsec:result}

In Table~\ref{table:recall}, the proposed mechanism achieved 100$\%$
and 78$\%$ Recall rate for the target Web applications. In
identifying vulnerabilities, our mutated XSS attacks found 27
vulnerabilities and missed 4 vulnerabilities. After the manual
investigation, the two of four missed vulnerabilities could be
ignored. Since input was sanitized by the PHP function, the ``html
specialchars'' before it was saved into the database. This
``htmlspecialchars'' function is used to convert ``\&'' (ampersand),
``$<$'', and ``$>$'' into HTML markup, (i.e., ``\&'' as ``\&amp;'',
``$<$'' as ``\&lt;'' and ``$>$'' as ``\&gt;''). When the converted
input was retrieved from the database, it was already harmless.
However, some un-converted part of input were still displayed on the
page title, it might be the reason of being falsely determined as
vulnerabilities. If adopting more strict test oracle, the false
positives would not be occurred. The remained two vulnerabilities
were the stored XSS vulnerabilities, and the attack, which was input
from administrative page, would be triggered when the clients
browsed the login page. The two vulnerabilities did not be sanitized
by any filter functions, so our mutated XSS attacks succeeded in
being saved into the database. \comment{It was pity that the system
did not check other Web pages if they were suffered from attacks, so
the system missed the two vulnerabilities. It is clearly that
generated mutated XSS attacks, which contain sufficient attack
vectors to penetrate the two Web applications.}

\begin{table}
\begin{center}
\caption[The Recall of our approach]{The Recall of our approach (The
higher Recall rate denotes that the set of mutated XSS attacks
really contain the manners to identify XSS vulnerabilities.)}
\label{table:recall}
\begin{tabular}{@{}cccccp{10cm}@{}}
\hline \textbf{Target}      & \textbf{Identified}      & \textbf{Missed}          & \textbf{Total}          & \textbf{Recall} \\
       \textbf{Application} & \textbf{Vulnerabilities} & \textbf{Vulnerabilities} & \textbf{Vulnerabilities} & \textbf{Rate}  \\
\hline
webchess   & 13 & 0 & 13 & 100$\%$ \\
schoolmate & 14 & 4 & 18 &  78$\%$ \\
\hline
\end{tabular}
\end{center}
\end{table}

In reality, the system was able to generate more than 1,500,000 XSS
attacks, since the referred XSS attack was with longer structure and
had more plain-text type of tokens. We limited the system generated
928 and 300 mutated XSS attacks respectively in evaluation of
testing target Web application.  The FP rate regarded to
``Webchess'' and ``schoolmate'' were 53.1$\%$ and 78.7$\%$
respectively, as shown in Table~\ref{table:fp}. The performance is
worth to mention that our proposed approach actually did not
consider any prior knowledge about target applications, our mutated
XSS attacks were still effective for testing the vulnerabilities.

\begin{table}
\begin{center}
\caption[The performance of our mutated XSS attacks]{The performance
of our mutated XSS attacks (The lower FP rate denotes that the set
of mutated XSS attacks were with more threats to the target Web
applications.)} \label{table:fp}
\begin{tabular}{@{}cccccp{12cm}@{}}
\hline \textbf{Target}      & \textbf{True Positive} & \textbf{False Positive} & \textbf{Reported Successful} & \textbf{False Positive} \\
       \textbf{Application} & \textbf{Attacks}       & \textbf{Attacks}        & \textbf{Attacks}             & \textbf{Rate}           \\
\hline
webchess   & 435 & 493 & 928 & 53.1$\%$ \\
schoolmate & 64  & 236 & 300 & 78.7$\%$ \\
\hline
\end{tabular}
\end{center}
\end{table}

In the aspect of manifestation of generated mutated XSS attacks, we
found several interesting attack vectors as shown in
Table~\ref{tabel:mutated samples}. The first attack--the attack
vector was "$>$ and the attack body was
$<$iframe/src=http://xssed.com$>$--was preferred to place between
the double quotes as the value of an attribute. According to W3C
specification, the tag "iframe" and the attribute "src" need to be
separated by a space, but the attack body adopted the slash instead
of the space and this format was recognized by Internet Explorer 8.
An attacker assigns most likely the malicious Web site as the value
of the attribute ``src'' to trigger drive-by download attack without
any knowledge of client if both of the size of width and the height
of iframe tag equal to zero. The drive-by download
attack~\cite{2007:ghost, 2008:iframe} is a serious security threat
in recent years, because this type of attacks could be downloaded
the spywares, computer viruses, like Trojan horses and bot programs,
by exploiting Web browser vulnerabilities. In the case of the second
attack, the attack vector was the same as the first one, but the
attack body showed the common attack behavior that invokes
JavaScript interpreter. It was worthwhile to notice the single input
tag in second attack that showed an input field in the page which
was rendered by the browsers of Microsoft Internet Explorer 8 or
Google Chrome. We thought that input field could be utilized to
entice clients into input personal confidential data through social
engineering, like phishing attacks.

The attack vector of the third attack was $>$"$>$ that could be
seamlessly integrated with the HTML page because the first $>$ was
as the value of an attribute, and in the meantime the second $>$ was
as the indicator of ending a tag. The combination of digital value
(or characteristics), value delimiter, and end-tag-indicator (e.g.,
1'$>$) is the popular style to make subsequent tags or attack bodies
be correctly rendered by Web browsers. For example, in the third
attack, the marquee tag and attack body (i.e.,
\emph{$<$script$>$alert(123)$<$/script$>$}) have effectiveness to
scroll the following texts (e.g., the reader images the "XSS is
here" instead of "alert(123)") and popup a dialogue window
respectively. The fourth attack string was different with the third
attack vector, and the difference was no prefixed characteristic or
value because the /$>$ was as end-tag-indicator. The fifth attack
string $>$"$>$ was as well as the third attack vector, but former
was suffixed with $<$/form$>$, that was often used in input field,
while inputs were output in a form block. In such situation,
attackers could close the original form block for creating another
form block to convince clients enter their own data and redirect the
input data.

\begin{table}
\begin{center}
\caption[The samples of mutated XSS attack]{The samples of mutated
XSS attack (The \emph{attack vectors} are separated by commas, and
the \emph{attack bodies} are denoted as
italic.)}\label{tabel:mutated samples}
\begin{tabular}{cl}
\hline
\multicolumn{2}{c}{\textbf{Mutated XSS Attack Samples}}\\
\hline
1. & {"$>$, alert(123)\emph{$<$iframe/src=http://xssed.com$>$}alert(123)$<$/scrihttp://pt$>$alert(123)}\\
2. & {"$>$, '$>$$<$/div$>$alert(123)$<$input$>$\emph{$<$script$>$alert(123)$<$/script$>$}$<$/marquee$>$alert(}\\&
{123)"$>$}\\
3. & {$>$"$>$, $<$/p$>$alert(123)$<$marquee$>$\emph{$<$script$>$alert(123)$<$/script$>$}$<$/title$>$alert(123)}\\
4. & {"/$>$, $<$/ScRiPt$>$alert(123)$<$title$>$\emph{$<$script$>$alert(123)$<$/script$>$}$<$/SCRIPT$>$alert(}\\&
{123)}\\
5. & {$>$"$>$, $<$/form$>$alert(123)$<$b$>$$<$script$>$alert(123)$<$/script$>$$<$/input$>$alert(123)" t}\\&{type="hidden" /$>$}\\
\hline
\end{tabular}
\end{center}
\end{table}
These results reveal that mutated XSS attacks can manipulate the face of response pages by HTML tags and trigger the JavaScript interpreter by inducing
additional script-inducing constructs. This is helpful to Web application programmers to know what combination
of HTML tags and special symbols they must handle carefully.

\subsection{Discussion}\label{subsec:analyze}

Our mutated XSS attacks can be used to test XSS blacklists in the
Web application. This type of sanitization functions define a set of
restricted XSS patterns for filtering the input. For example,
phyMyAdmin, a popular administration of MySQL database in Web
applications, had a blacklist bypass vulnerability in version 2.8.0
to 2.9.2. The cause resulted from ignoring the end tag,
$<$/SCRIPT$>$, and only checking the end tag, $<$/script$>$. This
vulnerability can break the section of JavaScript to interrupt
original processing. Hence, an attacker can send an attack, like
$<$/SCRIPT$>$$<$script$>$alert(123)$<$/script$>$ and invoke the
JavaScript interpreter for popping a warning window on the client's
browser. The attack can be generated by our proposed approach, and exists in XSS cheat sheet. It is clear that the structural
learning of attack vectors is efficient and benefited from the real
XSS exploits. On the other hand, we ever tested the public XSS
detection rule against the mutated XSS attacks, but the result
showed that all mutated XSS attacks were detected. Because the
detection rules filter all characters that range between a to z,
between A to Z and \% between $<$ and $>$, this may detect all XSS
attacks, but the false positives were still high. We believe that
the the testing ability of mutated XSS attacks generated by the
proposed approach is as well as the XSS cheat sheet made by experts.
\section{Related Work}\label{sec:past}
Generally, three techniques are used to identify the XSS
vulnerabilities in different locations and phases. These three
techniques include static analysis~\cite{2006:Pixy, 2008:StaticanalysisofXSS}, black-box testing~\cite{2010:StateoftheArt:blackbox, 2004BypassTesting} and hybrid
testing~\cite{2008:Saner, 2009:Ardilla}. Different techniques are
appropriate at different usages with their own characteristics (
strength and weakness).

Static analysis could
take the whole possible values of a variable into consideration based on the source code of the Web application without
executing the application. However, some fundamental problems result in more higher false positives and fail to prove the
feasibility of vulnerabilities.

Black-box approach detects vulnerabilities by sending inputs to an application
without priori knowledge about the target applications. Generally, the testing tools are equipped
with a lot of attack payloads for triggering specified vulnerability.
This type of testing not only mimics real-world attacks
from malicious users, but also provides cost-effective testing for
identifying a range of serious vulnerabilities.

SecuBat~\cite{2006:Secubat} utilized attack patterns
(e.g., SQL and XSS vulnerabilities) and injected them into entry points. Then the responses from the server were analyzed to
identify the vulnerabilities of the web application. However, SecuBat used fixed attack patterns for identifying SQL
injection or XSS vulnerabilities and ignored knowledge about the
filter functions of applications. On the other hand, our approach
learns evasion techniques from real-world attacks and they are able
to extend the ability of black-box testing.
McAllister et al.~\cite{2008:LeveragingUserInteractions} performed
in-depth testing of web applications against XSS attacks by
leveraging usage-based information, modifying them with attack test
cases, and replaying them back. Similar to
our approach, we collects attacks that are practical and
malicious, and generates attacks with malicious intension.

Hybrid testing combines static and
dynamic analysis to verify whether identified flaws are real
vulnerabilities by executing test data. In order to enlarge the scope of testing and
speed up the testing period, an automated attack-testing mechanism
is required.

The above works need to maintain a corpus of XSS attacks collected from
public sources or suggested by experts. These XSS attacks are examined for
their attack specifications or patterns to identify XSS vulnerability.
However, attacks generated in such way did not benefit from the
sources, for example, the structures or elements of XSS attacks. To implement an attack,
it must be crafted specifically for a certain weakness of a web
application. 
\section{Conclusion}\label{sec:conc}
We have proposed a technique to automatically generate cross-site
scripting (XSS) attacks for identifying XSS vulnerabilities in Web
applications. For simulating the manner of crafting attacks used by
attackers, this technique takes the archives of XSS attack as input,
and is based on hidden Markov model (HMM) as well as Bayes'
Theorem for learning the general structure of the attack vectors.
Thus, generation mechanism takes the advantage of structure model to
produce mutated XSS attacks like the behavior of the attackers. In
our experiments, the technique effectively produces the XSS
attacks that trigger XSS vulnerabilities in target applications, and
gives a help to provide a set of various mutated XSS attacks, which are essentially taken into consideration by web
programmers.

\section*{Acknowledgements}
This work was supported in part by Taiwan Information Security Center at
NTUST(TWISC@NTUST), National Science Council under the grants NSC 99-2219-E-011-004, and by the National Science Council of Taiwan under
grants 96-2221-E-011-064-MY3. The authors also gratefully acknowledge the helpful
comments and suggestions of the reviewers, which have improved the
presentation.

\bibliographystyle{eptcs} 
\bibliography{bib/my}
\end{document}